\documentclass[fullpaper]{nldl}

\renewcommand{\DOI}{12345}

\usepackage[utf8]{inputenc}
\usepackage{url}
\usepackage{graphicx}
\usepackage{authblk}
\usepackage{multirow}

\usepackage{algorithm}
\usepackage{algorithmic}
\usepackage{amsmath}
\usepackage{balance}

\usepackage{hyperref}

\title{Fast accuracy estimation of deep learning based multi-class musical source separation}
\author[1, 2]{Alexandru Mocanu\thanks{Work completed during an internship at Logitech.}}
\author[1, 3]{Benjamin Ricaud}
\author[2]{Milos Cernak}
\affil[1]{LTS2, \'{E}cole Polytechnique F\'{e}d\'{e}rale de Lausanne, Station 11, CH-1015 Lausanne, Switzerland}
\affil[2]{Logitech Europe S.A., CH-1015, Lausanne, Switzerland}
\affil[3]{Department of Physics and Technology, UiT The Arctic University of Norway, Tromsø NO-9037, Norway}

\date{\vspace{-5ex}}

\begin{document}
\nldlmaketitle

\begin{abstract}  
Music source separation represents the task of extracting all the instruments from a given song. Recent breakthroughs on this challenge have gravitated around a single dataset, MUSDB, only limited to four instrument classes. Larger datasets and more instruments are costly and time-consuming in collecting data and training deep neural networks (DNNs). In this work,  we propose a fast method to evaluate the separability of instruments in any dataset without training and tuning a DNN.
This separability measure helps to select appropriate samples for the efficient training of neural networks. Based on the oracle principle with an ideal ratio mask, our approach is an excellent proxy to estimate the separation performances of state-of-the-art deep learning approaches such as TasNet or Open-Unmix.
Our results contribute to revealing two essential points for audio source separation: 1) the ideal ratio mask, although light and straightforward, provides an accurate measure of the audio separability performance of recent neural nets, and 2) new end-to-end learning methods such as Tasnet, that operate directly on waveforms, are, in fact, internally building a Time-Frequency (TF) representation, so that they encounter the same limitations as the TF based-methods when separating audio pattern overlapping in the TF plane. 
\end{abstract}

\section{Introduction}
\label{sec:intro}

Music, along with speech, is a main point of interest for the field of audio processing. The extraction of individual instrument tracks from a song has been of great interest for a couple of decades. Most of the initial efforts were based on signal processing techniques and used unsupervised methods to exploit signal characteristics such as sparsity~\cite{plumbley2009sparse}, harmonicity and inharmonicity \cite{rigaud2013does}, harmonic and percussive spectra \cite{laroche2015structured, fitzgerald2010harmonic}, beat \cite{kameoka2012constrained} or repetitiveness \cite{rafii2011simple, fitzgerald2012vocal}. However, these methods do not separate the instruments but rather some sound patterns produced by them.
With the rise of deep learning, the latest models~\cite{luo2019conv,defossez2019music} provide a better separation quality and manage to extract the instruments with all their sound specificities. Nevertheless, a substantial inconvenience for the deep learning models comes from their supervised learning training procedure that requests large amounts of labeled data, namely songs and their separate instrument tracks. 

The most widespread dataset used for training deep learning models for music source separation, MUSDB18 \cite{musdb18}, only provides four instrument classes (vocals, drums, bass, and others). This is a significant limitation for the community. Creating a new dataset is time-consuming and may have a high cost. Moreover, after spending much time training deep networks on it, it may be inappropriate to train particular classes of instruments. 

The proposed method is based on the Ideal Ratio Mask (IRM) oracle \cite{stoter20182018} and does not require any training. It outputs a separation score for each instrument class of the input tracks from a song and its separated track.

Our contribution is twofold. Firstly, we show that the oracle is a good indicator for source separation performance. We do this by comparing the separation score of the oracle to that of a trained proof-of-concept TasNet-based model. The results of the two models are strongly correlated. We tried the separation on electric guitar, which overlaps in the time-frequency domain with bass and vocals. It confirms that the separation of new instrument classes (beyond the MUSDB18 4 classes) can be obtained successfully with deep learning models.
Secondly, by using a TasNet-based model and the IRM oracle, we show a connection between waverform-based and spectrogram-based models. The IRM looks at separability in terms of spectro-temporal overlap, and the Tasnet-based models operate directly on the waveform. In this context, a correlation between the two seemed much less trivial. There seems to be a more interesting relationship that indicates that even if TasNet is end-to-end learning, internally, it is a TF-based approach. Thus it cannot separate well sounds that are overlapping in the TF plane.

The structure of the paper is as follows: Section~\ref{sec:related_work} introduces related work. Section~\ref{sec:choose_instruments} describes our proposed method, and Section~\ref{sec:experiments} presents the experimental setup and obtained results. Section~\ref{sec:conclusion} concludes the paper.

\section{Related work}
\label{sec:related_work}

Deep learning methods have been prevalent in music source separation in the last few years. Most of the models separate only the four classes provided in the MUSDB \cite{musdb18} dataset. Takahashi et al. \cite{takahashi2018mmdenselstm} combine convolutional dense blocks and Long short-term memory blocks and apply them on multiple frequency bands to capture features specific to each band. This model achieved the best performance in the SiSEC 2018 \cite{stoter20182018} competition but used additional proprietary training data. St{\"o}ter et al. \cite{stoter2019open} proposed a recurrent model that trains a separate network for each instrument. They made the code open source and trained their model only on MUSDB to serve as a benchmark for future research. 

Many deep learning models operate on the magnitude spectrogram, as this provides a representation that is both intuitive and has proven helpful in several audio processing tasks. End-to-end systems, which also learn how to encode and decode the raw signal, have also attracted more attention. While Stoller et al. \cite{stoller2018wave}, who adapted a U-Net architecture, did not manage to outperform the best spectrogram-based models, Samuel et al. \cite{samuel2020meta} managed to reach state-of-the-art performance. Their model reuses the temporal convolutional network from Conv-TasNet \cite{luo2019conv}, for which it learns to generate parameters based on instrument embeddings. It also includes a complex encoder that leverages convolutions and spectrograms, and it operates in a multi-stage fashion, receiving as input mixture signals resampled at three different sampling rates. Deffosez et al. \cite{defossez2019music} also adapt the Conv-TasNet architecture and obtain even better performance than Meta-TasNet, a model with more parameters. These recent increases in performances are at the cost of more complex architectures that require more computing power.

There have also been a few attempts at separating music into more classes. Hennequin et al. \cite{hennequin2020spleeter} use a U-Net \cite{jansson2017singing} architecture and add the piano class as well. However, they do not release the results for piano separation. Hung and Lerch \cite{hung2020multitask} propose a multitask learning solution. They also detect instrument activity, extend the number of instruments to six, and add besides the piano, electric, and acoustic guitars. They use open-source data to train their models. Nevertheless, they use three large datasets of songs together with data augmentation, which requires more computing power. 

\section{Method}
\label{sec:choose_instruments}



\begin{figure*}[htb]
    \centering
    \includegraphics[width=0.8\linewidth]{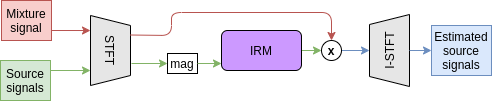}
    \caption{Oracle source separator based on Ideal Ratio Mask. The TF representations of both the mixture and source signals are computed using the Short Time Fourier Transform (STFT). Then, masks are applied from the magnitude (mag) of the source signals. The masked signals are represented back in the time domain with an Inverse STFT (I-STFT).}
    \label{fig:irm_separator}
\end{figure*}

The IRM oracle \cite{stoter20182018} is typically used as a baseline when evaluating the performance of a source separation model. It acts as an upper bound for spectrogram-based models that apply masks on the spectrogram to perform the separation. The IRM is defined as follows: given the spectrograms $y_j(f, t, i)$ for source signals $s_j$, $1 \le j \le J$, where $J$ is the number of sources, and $(f, t)$ are the TF bin coordinates for channel $i$, the IRM separation mask for source $j$ is computed as:
\begin{equation}
    M_j(f, t, i) = \frac{v_j(f, t, i)}{\sum_{j'=1}^J v_{j'}(f, t, i)}
\end{equation}
where $v_j(f, t, i) = |y_j(f, t, i)|^{\alpha}$, $\alpha \ge 0$. In our case, we choose $\alpha = 2$ to use power spectrograms, which usually give the best performing oracle. Figure \ref{fig:irm_separator} depicts the structure of the oracle source separator. For a given song, we compute the spectrograms for both the mixture signal and the source signals. Then, we take the magnitude spectrograms for the sources and feed them to the IRM that outputs masks for each source. After that, the masks are applied to the mixture spectrogram, and finally, the estimated source spectrograms are converted into source signal estimates through an inverse short-time Fourier transform.

To test a configuration of instrument classes, we use the IRM oracle for performing the separation. Then we evaluate the following metrics:
\begin{flalign}
\text{SDR} &= 10 \log_{10} \frac{\|s\|^2}{\|e_{spat} + e_{interf} + e_{artif}\|^2} \\
\text{ISR} &= 10 \log_{10} \frac{\|s\|^2}{\|e_{spat}\|^2} \\
\text{SIR} &= 10 \log_{10} \frac{\|s + e_{spat}\|^2}{\|e_{intef}\|^2} \\
\text{SAR} &= 10 \log_{10} \frac{\|s + e_{spat} + e_{interf}\|^2}{\|e_{artif}\|^2} \\
\text{SI-SDR} &= 10 \log_{10} \frac{\|\alpha s\|^2}{\|\alpha s - \hat{s}\|^2}, \alpha = \frac{\hat{s}^T s}{\|s\|^2}
\end{flalign}
where $s$ is the clean source signal, $\hat{s}$ is the estimated source signal given by $\hat{s} = s + e_{spat} + e_{interf} + e_{artif}$ with $e_{spat}$ the error due to spatial distortions, $e_{inter}$ the error due to interference with other sources and $e_{artif}$ the error due to artifacts. The Source to Distortion Ratio (SDR), Image to Spatial Distortion Ratio (ISR), Source to Interference Ratio (SIR), and Source to Artifacts Ratio (SAR) are computed using the \textit{museval} Python package.
In contrast, the Scale-Invariant SDR (SI-SDR)~\cite{le2019sdr} is computed in the same way as the other metrics from \textit{museval}, by first computing for each song the median score over one-second windows and then taking the median over the songs. The primary metric of interest is usually the SDR. However, as the relative scale of the real and estimated source signals influences its value, we examine the SI-SDR when deciding what instrument classes to choose.

\section{Experiments}
\label{sec:experiments}

\begin{table}
    \centering
    \begin{tabular}{clccccc}
        \hline
        & & SI-SDR & SDR & SIR & ISR & SAR \\
        \hline
        \parbox[t]{2mm}{\multirow{6}{*}{\rotatebox[origin=c]{90}{IRM}}} & vocals & 8.43 & 9.02 & 17.42 & 16.38 & 9.26 \\
        & drums & 5.11 & 6.24 & 14.40 & 10.31 & 5.74 \\
        & bass & 9.55 & 9.99 & 14.09 & 15.38 & 11.44 \\
        & eguitar & 2.83 & 4.62 & 10.61 & 9.08 & 4.86 \\
        & wind & -2.76 & 1.06 & -9.48 & 7.81 & 0.11 \\
        & other & 3.45 & 5.04 & 10.43 & 9.24 & 5.24 \\
        \hline
        \parbox[t]{2mm}{\multirow{6}{*}{\rotatebox[origin=c]{90}{Meta-TasNet}}} & vocals & 1.36 & 3.52 & 8.27 & 6.81 & 3.74 \\
        & drums & 1.36 & 3.62 & 15.39 & 5.30 & 3.88 \\
        & bass & 9.00 & 9.42 & 14.17 & 13.56 & 10.35 \\
        & eguitar & -6.82 & 0.46 & -2.83 & 2.73 & 2.11 \\
        & wind & -34.55 & -11.22 & -18.90 & 0.38 & 0.66 \\
        & other & -3.77 & 0.92 & -1.12 & 4.75 & 3.49 \\
        \hline
    \end{tabular}
    \caption{IRM and Meta-TasNet scores on Karaoke Version dataset.}
    \label{tab:karaoke}
\end{table}

\subsection{Experimental Setup}

We use our custom dataset for training and testing, obtained from downloading songs from the Karaoke Version website\footnote{\url{https://www.karaoke-version.com}}. We use 241 train, 46 validation, and 50 test songs sampled at 44.1 kHz, totaling 23 hours of music. We also experiment with training on smaller subsets of 24 songs. We included the 4 MUSDB classes for the experiments and added two new ones: electric guitar and wind instruments (like saxophone or flute). We generated equally loud source tracks for vocals, drums, bass, electric guitar, wind, and other instrument classes by mixing instrument signals appropriately for each of the songs.

To validate the IRM oracle's role as an indicator for the performance of a deep learning model, we train a variation of Meta-TasNet\footnote{\url{https://github.com/pfnet-research/meta-tasnet}} described in~\cite{samuel2020meta}. The original model consists of three stages: separate signals resampled at 8, 16, and 32 kHz and reuse information from the previous stage. We only keep the first stage, training a single network that separates signals resampled at 8 kHz. The rest of the network is left unchanged. At evaluation time, the separated 8 kHz signals are upsampled to 44.1 kHz, the original sampling rate of the data. We train our 8kHz-Meta-TasNet model for 80 epochs. The separation performances are lower than the full Meta-TasNet at 44.1 kHz (due to smaller network and downsampling to 8kHz, which lose the high-frequency components) but are still decent. It permits to obtain a good compromise between the network accuracy and computing time. Our goal is to estimate the ability to separate instruments and not to obtain the best separation possible.


\subsection{Results}

Table \ref{tab:karaoke} shows the results of the IRM oracle applied to the Karaoke Version test set. We notice that the initial classes, vocals, drums, and bass, yield the best scores. However, compared to the results obtained on MUSDB \cite{musdb18}, drums obtain an SI-SDR score of more than 3 dB lower. We still expect drums to be quite well separated on this dataset, so we assume we can separate instruments with scores that may be as low as about 2.5 dB. In this case, we predict that we can have decent separation for all classes except wind. In the wind instruments class, the problem originates from the severe interference with other instruments, as shown by the SIR score.

After running the Meta-TasNet model on the test set, we obtained the bottom of Table \ref{tab:karaoke}. Given the small size of the model, the scores are lower than for the full-size state-of-the-art model but still provide intuition for what would happen when scaling up the model size. For example, we notice that vocals, drums, and bass do indeed maintain excellent scores even in these conditions. In contrast, electric guitar and other classes still do better than having noise as loud as the clean signal, i.e., having an SDR score of 0 dB. On the other hand, the wind instruments suffer from interference that drags down the SDR score, just as predicted by IRM. Therefore, IRM indicates which instrument classes would have good or bad separation scores when using deep learning approaches.

\begin{table}
    \centering
    \begin{tabular}{clccccc}
        \hline
        & & SI-SDR & SDR & SIR & ISR & SAR \\
        \hline
        \parbox[t]{2mm}{\multirow{6}{*}{\rotatebox[origin=c]{90}{Pearson}}} & vocals & 0.96 & 0.81 & 0.89 & 0.86 & 0.84 \\
        & drums & 0.46 & 0.56 & 0.88 & 0.91 & 0.81 \\
        & bass & 0.87 & 0.68 & 0.88 & 0.82 & 0.72 \\
        & eguitar & 0.88 & 0.67 & 0.90 & 0.74 & 0.60 \\
        & wind & 0.80 & 0.76 & 0.87 & 0.81 & 0.40 \\
        & other & 0.71 & 0.47 & 0.66 & 0.57 & 0.61 \\
        \hline
        \parbox[t]{2mm}{\multirow{6}{*}{\rotatebox[origin=c]{90}{Spearman}}} & vocals & 0.84 & 0.83 & 0.53 & 0.54 & 0.87 \\
        & drums & 0.78 & 0.76 & 0.67 & 0.73 & 0.80 \\
        & bass & 0.82 & 0.82 & 0.80 & 0.42 & 0.75 \\
        & eguitar & 0.76 & 0.78 & 0.49 & 0.41 & 0.56 \\
        & wind & 0.75 & 0.89 & 0.84 & 0.82 & 0.36 \\
        & other & 0.69 & 0.57 & 0.37 & 0.51 & 0.49 \\
        \hline
    \end{tabular}
    \caption{Correlations between IRM and Meta-TasNet scores.}
    \label{tab:correlation}
\end{table}

We computed the Pearson and Spearman correlations of the two models, for each metric, on the scores obtained \emph{on every individual song} from the test set. Table~\ref{tab:correlation} shows these correlations between the results of IRM and Meta-TasNet on the 50 random test tracks for each of the instrument classes. In both cases, the correlations that we obtain for most of the metrics are strong (above 0.6) or very strong (above 0.8). This confirms that IRM serves as a good indicator for source separation performance on the dataset.

\begin{figure}
    \centering
    \includegraphics[width=\columnwidth]{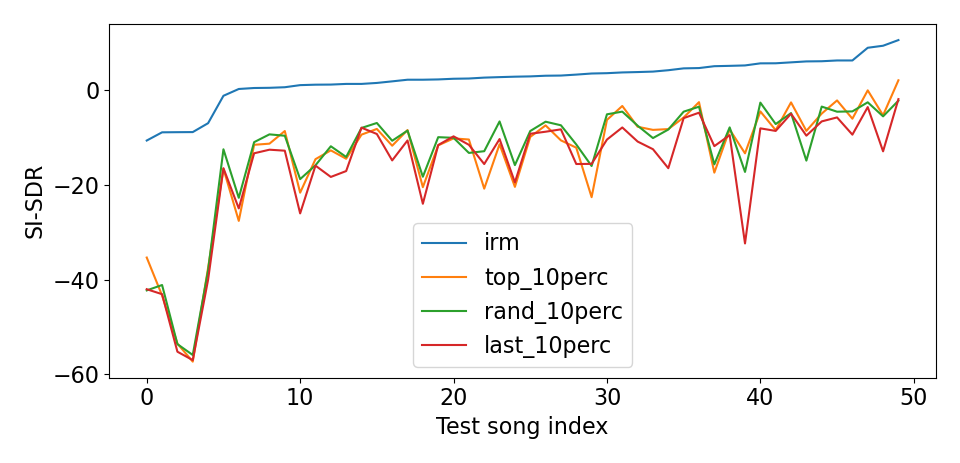}
    \caption{SI-SDR scores for electric guitar on the test set sorted by IRM performance for Meta-TasNet models trained on subsets of the training set.}
    \label{fig:meta_subsample}
\end{figure}

To further confirm the relationship between the scores provided by the IRM oracle and those obtained by a Meta-TasNet model, we trained on samples consisting of 10\% of the songs of the initial training set. We selected samples by three different criteria: the ``best`` songs (top\_10perc), random songs (rand\_10perc), the ``worst`` songs (last\_10perc) according to the SI-SDR score of the IRM model. Figure \ref{fig:meta_subsample} presents the SI-SDR scores for electric guitar on the test set for the IRM model and the Meta-TasNet model trained with the three different samples. The songs are sorted by the SI-SDR score of the IRM model. We see that no matter what data we train on, the results of the models still follow the trend predicted by the IRM model. Table \ref{tab:subsample_correlations} shows the correlations between the Meta-TasNet results and the IRM results, solidifying our statement. We can also note from Fig.~\ref{fig:meta_subsample} the close separability scores obtained from the different training subsets. It highlights the relative stability of the separation process when training on different training sets, with different time-frequency separabilities. We notice however, on average, a slightly smaller SI-SDR score for the neural network trained on the worst IRM separable samples (red curve).

An interesting aspect to note is that IRM is an oracle that uses a TF representation to compute its metrics, while the Meta-TasNet model learns directly from waveforms. This implies that IRM acts as a universal predictor, being useful for both TF-based and waveform-based models.

\begin{table}
    \centering
    \begin{tabular}{lcc}
        \hline
         & Pearson & Spearman \\
        \hline
        top\_10perc & 0.88 & 0.76 \\
        rand\_10perc & 0.88 & 0.70 \\
        last\_10perc & 0.85 & 0.69 \\
        \hline
    \end{tabular}
    \caption{Correlations between IRM and Meta-TasNet SI-SDR scores on electric guitar.}
    \label{tab:subsample_correlations}
\end{table}

\begin{figure}
    \centering
    \includegraphics[width=\columnwidth]{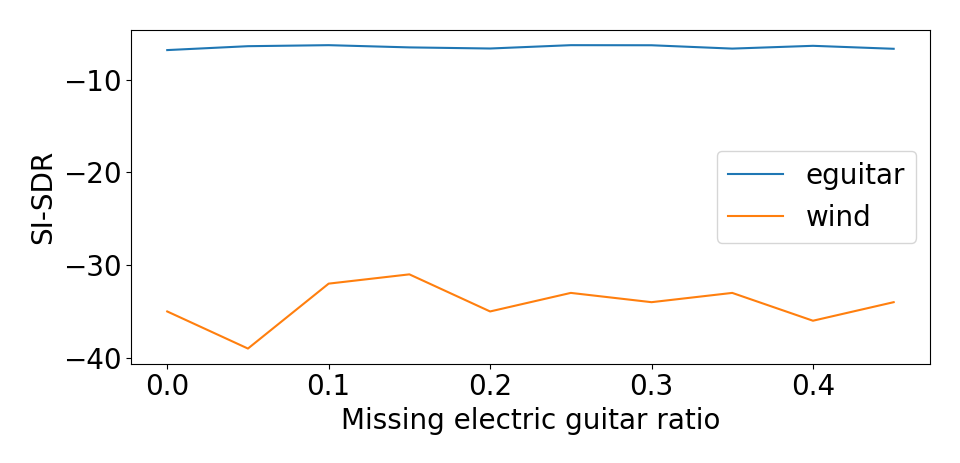}
    \caption{Meta-TasNet results for different ratios of muted electric guitar and wind instruments.}
    \label{fig:ratio_results}
\end{figure}

Hung and Lerch~\cite{hung2020multitask} choose the instrument categories based on their frequency of appearance in the dataset. A natural question is whether the IRM oracle would be a better indicator of the separation performance than the simple frequency of an instrument in the dataset. To analyze this situation, we took the training set and muted some electric guitar tracks. The test set was kept intact. We then trained our Meta-TasNet model on this data. We repeated this process for ratios of muted tracks ranging from 0.00 to 0.45 with a step of 0.05. 
Fig.~\ref{fig:ratio_results} displays the SI-SDR score of the Meta-TasNet model on the electric guitar and wind instrument classes. We see that even when muting the electric guitar considerably, the Meta-TasNet performance stays stable. The electric guitar scores maintain a large margin of about 28 dB compared to the wind instruments. 
For the wind instruments, no drop in performance was expected, as no tracks were muted. Since the IRM oracle is not a learning process, its performance is independent of the number of samples used in the training set. We thus confirmed that the IRM oracle is a much better indicator of separation performance than the frequency of an instrument in the dataset. This conclusion is valid for the Meta-TasNet architecture and within a reasonable minimal number of samples for each instrument category.

\section{Conclusion}
\label{sec:conclusion}

Increasing the size of the dataset or the network often leads to better performance for deep learning models. This is the actual trend, which implies increasing computing resources not everyone can afford. 
We propose a new way for evaluating the separability of audio sources in recordings, prior to the training of any neural network. 
To assess the quality of samples, we take advantage of the TF masking approach and the possibility to make an oracle in this particular configuration.

We reveal the close relationship between mask-based learning approaches such as Meta-TasNet and the TF representation. 
Our results indicate that even if TasNet has the freedom to learn a latent space where instruments would be separated efficiently by a masking process, this space is no better than the TF representation.
Hence, in order to improve the separation of sources overlapping in the TF plane, new architectures are needed.


\balance
\bibliographystyle{abbrv}
\bibliography{references}

\end{document}